\documentclass[amsfonts, amssymb, prl, superscriptaddress, notitlepage, twocolumn, nofootinbib]{revtex4-2}

\usepackage{amsmath}    
\usepackage{graphicx}   
\usepackage{verbatim}   
\usepackage{color}      
\usepackage{hyperref}   
\usepackage[english]{babel}
\usepackage{ucs}
\usepackage[utf8x]{inputenc}
\usepackage{subfigure}
\usepackage[autostyle]{csquotes}
\usepackage{float}
\usepackage{textcomp}
\usepackage{xspace}
\usepackage{orcidlink}
\usepackage{tikz}

\newcommand{\bc}[2]{$\beta$#1cell#2}

\newcommand{\Ca}{Ca$^{2+}$\xspace}

\begin{document}

\title{Critical transitions in pancreatic islets}

\author{D.~Korošak \orcidlink{0000-0003-3818-1233} }
\thanks{Corresponding author:\\dean.korosak@um.si}
\affiliation{University of Maribor, Faculty of Medicine, Institute for Physiology, Maribor, Slovenia}
\affiliation{University of Maribor, Faculty of Civil Engineering, Transportation Engineering and Architecture, Maribor, Slovenia}

\author{S.~Postić}
\affiliation{Medical University of Vienna, Center for physiology and pharmacology, Vienna, Austria}

\author{A.~Stožer}
\affiliation{University of Maribor, Faculty of Medicine, Institute for Physiology, Maribor, Slovenia}

\author{B.~Podobnik}
\affiliation{University of Rijeka, Faculty of Civil Engineering, Rijeka, Croatia}

\author{M.~Slak Rupnik \orcidlink{0000-0002-3744-4882} }
\thanks{Corresponding author:\\marjan.slakrupnik@meduniwien.ac.at}
\affiliation{Medical University of Vienna, Center for physiology and pharmacology, Vienna, Austria}
\affiliation{University of Maribor, Faculty of Medicine, Institute for Physiology, Maribor, Slovenia}
\affiliation{Alma Mater Europaea University - European Center Maribor, Maribor, Slovenia}

\date{\today}

\begin{abstract}
Calcium signals in pancreatic \bc{ }{s} collectives show a sharp transition from 
uncorrelated to correlated state resembling a phase transition as the slowly increasing glucose concentration crosses the tipping point. 
However, the exact nature or the order of this phase transition is not well understood. 
Using confocal microscopy to record the collective calcium activation of \bc{ }{s} in an intact islet under 
changing glucose concentration 
in increasing and then decreasing way, we first show that in addition to the sharp transition, 
the coordinated calcium response exhibits a hysteresis indicating a critical, first order transition.
A network model of \bc{ }{s} combining link selection and coordination mechanisms capture the observed hysteresis loop and the 
critical nature of the transition. Our results point towards the understanding the role of islets as 
tipping elements in the pancreas that interconnected by 
perfusion, diffusion and innervation cause the tipping dynamics and abrupt insulin release. 
\end{abstract}

\maketitle 


The bio-sociological organization of \bc{ }{s} in pancreatic islets~\cite{pipeleers1987biosociology} plays a vital role in managing insulin release to ensure optimal nutrient distribution and metabolic balance in the body. Secretion of insulin from pancreas is biphasic~\cite{curry1968dynamics,nesher1987biphasic} with an immediate, sharp initial peak followed by a slower second phase. The fast first phase indicates that the response to stimulus from islets is essentially simultaneous with a precise and coherent \bc{ }{s} activation. The calcium activity of \bc{ }{s} at the level of individual islets, a key signal~\cite{wollheim1981regulation} in insulin secretion pathway, remains low and stochastic at lower glucose concentrations approximately around 6 mM. Once the gradually increasing glucose level hits the tipping point (around 7 mM), an abrupt shift in the correlation patterns and synchronicity of calcium activity among localized functional groups of \bc{ }{s} is experimentally observed~\cite{fernandez2000synchronous,stovzer2013glucose,stovzer2021glucose}. Such sudden change in a macroscopic state of islets~\cite{valdeolmillos1996vivo} from an incoherent into a synchronous functional state has considerable implications for insulin secretion and overall metabolic regulation, and is an example of a phase (critical) transition in a system of cells that interact mostly locally. 

The phase transition in \bc{ }{s} activity is a complex phenomenon that involves the interplay between glucose metabolism, electrical activity, and intracellular \Ca dynamics. The transition is also influenced by the spatial organization of \bc{ }{s} within the islet, the heterogeneity of \bc{ }{s} responses to glucose, and the strength of intercellular coupling. The efforts to understand the occurence of phase transitions in \bc{ }{s} activity have been made using a variety of experimental and theoretical approaches, often employing network representations of \bc{ }{s} interactions.
In one such study~\cite{hraha2014phase}, the authors showed that a sharp transition between active and inactive states in pancreatic islets occurs when the proportion of quiescent \bc{ }{s} exceeds some critical threshold. Using a dynamical model, the study suggests that the critical behavior of pancreatic islets follows general properties of coupled heterogeneous networks. Another study~\cite{stamper2014phase} showed that the onset of islet disfunction can be understood as a phase transition in the pancreatic islet \bc{ }{s} network. Using percolation theory, the authors demonstrated that there exists a critical threshold of \bc{ }{s} loss beyond which the cells fail to synchronize, and that both site occupancy and bond strength (intercellular coupling) are critical to ensure a proper islet function. 

Less attention was given to the nature of such phase transitions in \bc{ }{s} activity, i.e., whether the transition is of the first or the second order, or in other words whether the order parameter that describes the macroscopic state of the system changes continuously or discontinuously when the control parameter crosses the tipping point~\cite{stanley1971phase}. Percolation in random networks where new links are added randomly is a second order phase transition with the giant component appearing smoothly at a critical probability of links in the network. In contrast, explosive percolation~\cite{achlioptas2009explosive,nagler2011impact} is a sharp second order or even first order phase transition where the giant component appears abruptly at a critical number of links. 

In this work, we aim to address this question by deliberately crossing the tipping point in \bc{ }{s} activity by slowly increasing and then slowly decreasing the glucose concentration in intact pancreatic islet in fresh tissues slices. We show that the transition in \bc{ }{s} activity exhibits a hysteresis, indicating a critical, first order transition. 

We start by first displaying the data compiled from published experimental evidence for the sharp transition in \bc{ }{s} activity as a function of the control parameter in the upper panel of Figure 1. Active phase was defined either as plasma membrane electrical activity~\cite{sanchez1995electrical} or as intracellular \Ca  oscillations~\cite{benninger2011gap,hraha2014phase} of \bc{ }{s}. The dimensionless control parameter here is either the normalized glucose concentration~\cite{sanchez1995electrical, benninger2011gap}, or the normalized number of excitable cells in the population~\cite{hraha2014phase}.

\begin{figure}
\centering
\includegraphics[width=0.75\linewidth]{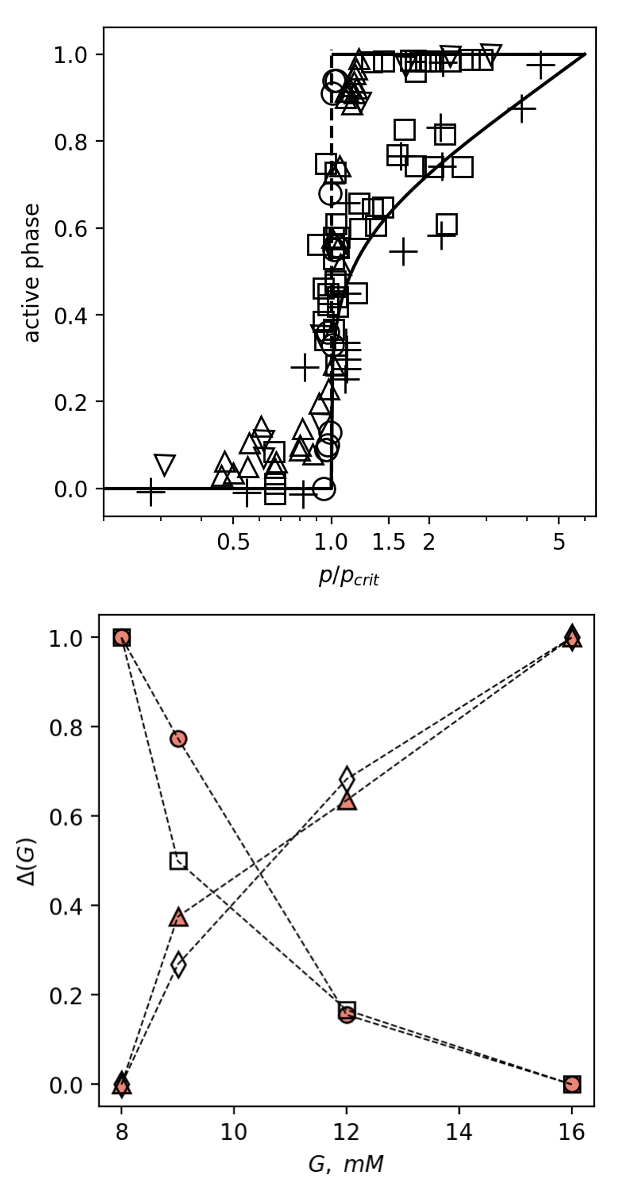}
\caption{\label{fig:fig1} {\bf Upper panel: Phase transition of \bc{ }{s} activity.} Active phase of \bc{ }{s} as a function of control parameter $p$. \bc{ }{s} electrical activity measured in islets in vivo (open squares), and from micro-dissected islets (pluses)~\cite{sanchez1995electrical} against scaled glucose concentration as a control parameter. [Ca$^{2+}$]$_{c}$ activity as a function of glucose concentration in fresh tissue slices (own data, open circles), and in \bc{ }{s} with Kir6.2 mutation (downward triangles)~\cite{benninger2011gap}. \bc{ }{s} electrical activity as a function of a scaled Kir6.2 mutation extent as coupling parameter (upward triangles)~\cite{hraha2014phase}. Data points shown here were extracted using WebPlotDigitizer~\cite{Rohatgi2020} from plots in original publications. 
{\bf Lower panel: Activation and deactivation delay of \bc{-}{s} }. Open symbols -- model, full symbols -- data~\cite{stovzer2013glucose}, circles, squares -- activation, triangles, diamonds -- deactivation. Here, $\Delta$ is the normalized difference $\Delta = (T_{d}-T_{d,min})/(T_{d,max}-T_{d,min})$ in activation/deactivation delays over the stimulation interval. 
}
\end{figure}

The observed sharp, first-order-like phase transition from inactive to active \bc{ }{s} states occurs at a particular critical value of the control parameter for each dataset and has practically the same shape for all data when the control parameter is scaled with its critical value. For a subgroup of cells, mostly from micro-dissected islet with impaired intercellular communication, we observe a smooth, second-order-like phase transition. 

The lower panel of Figure 1 shows the activation and deactivation delays $T_d$ of \bc{ }{s} as a function of glucose concentration~\cite{stovzer2021glucose}. The activation delay is the time it takes for the \Ca activity to reach a certain threshold after the glucose crosses the tipping point, while the deactivation delay is the time it takes for the \Ca activity to return to the baseline level after glucose drops below the tipping point. The activation and deactivation delays are normalized to the maximum and minimum values of the deactivation delay over the stimulation interval. The empirical data~\cite{stovzer2021glucose} was obtained from confocal microscopy recordings of \Ca activity in \bc{ }{s} in intact islets under the stimulation protocol where glucose increases with 1 mM/min from 6 mM to final concentration (8 mM, 12 mM, 14 mM, 16 mM) and decreases at the same rate. The model for the delays employs exponential increase and decrease of glucose concentration in the protocol and assumes constant rates and tipping points for activation and deactivation set at $G_c=7$ mM and $G_c=6.6$ mM (see Figure 2, top panel) independent of glucose.   

The combination of these empirical findings, the sharp transition in \bc{ }{s} activity and difference in the tipping points for activation/deactivation, indicate a critical, first-order phase transition in \bc{ }{s} activity. 
To further explore the nature of observed phase transition observed in \Ca activity data, we devised and performed a new experiment in which we deliberately first slowly increased and then slowly decreased the glucose concentration at the rate 1mM/min in intact pancreatic islet across the \bc{ }{s} stimulation threshold while simultaneously recording the \Ca activity. The experimental procedure, data acquisition and data analytics was, apart from the specifically shallow glucose ramp to capture critical activity threshold, the same as in our previous experimental work and described in details in~\cite{postic2023high}. The idea for this experiment was to capture crossing the tipping point and the subsequent return to the baseline state in a single experiment without the prolonged exposure to stimulating glucose levels. 

\begin{figure}
\centering
\includegraphics[width=1.0\linewidth]{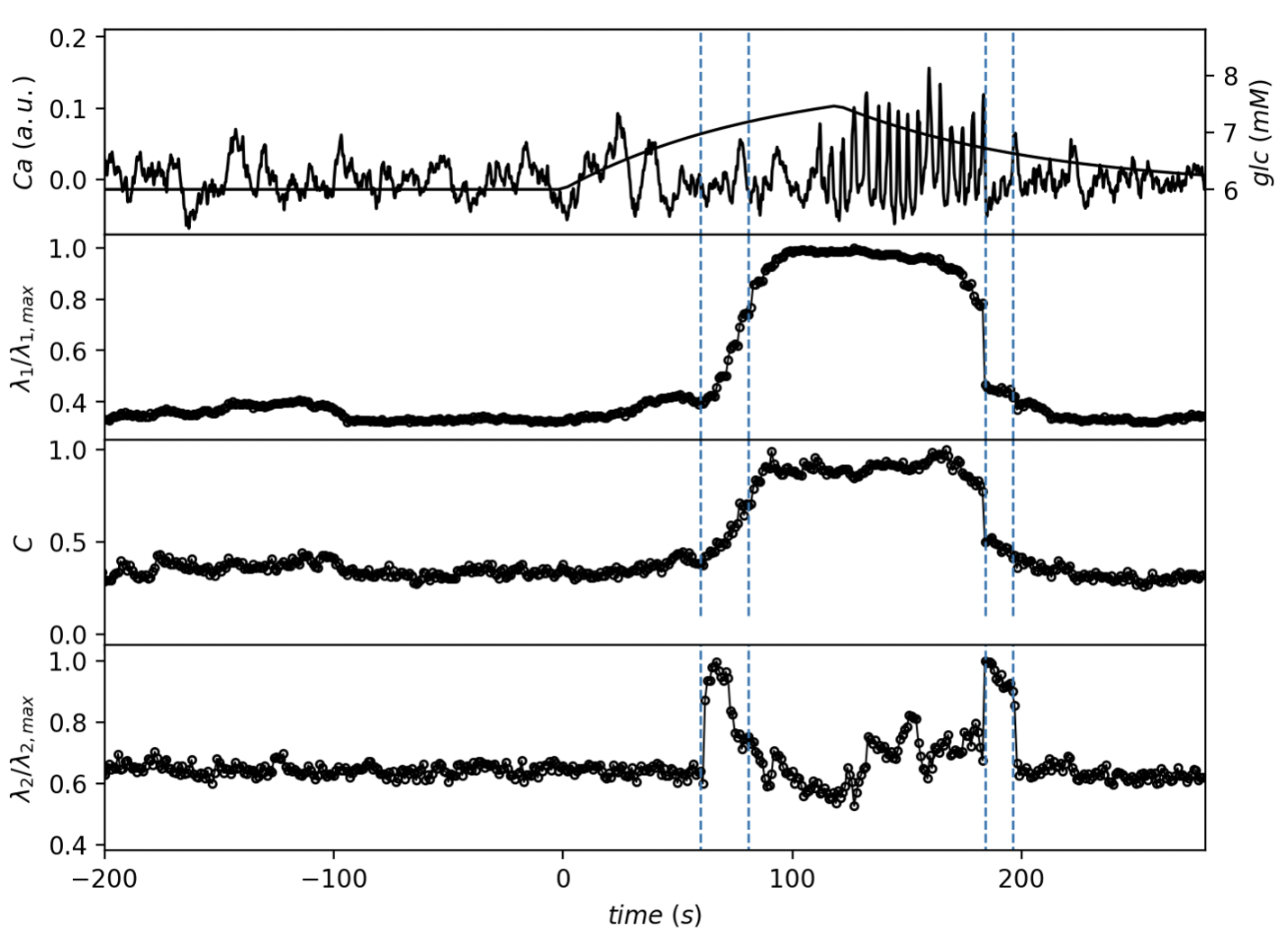}
\caption{\label{fig:fig2} {\bf Mean calcium signal timeseries and \bc{ }{s} correlation network properties.} 
Glucose 6-8 increase at a rate 1 mM/min. top to bottom: mean Ca signal, largest eigenvalue, 
clustering and second largest eigenvalue of the correlation network with fixed number of links. All networks were constructed with fixed mean 
degree $\langle k\rangle =10$.
}
\end{figure}

The top panel of Figure 2 shows measured mean Ca signal over all recorded \bc{ }{s} signals that shows the expected sharp transition from inactive to active state of \bc{ }{s} as the glucose concentration crosses the tipping point. The vertical dashed lines (first and third from left) mark the time points when the glucose concentration, shown as smooth full line, reaches the tipping point in the increasing part $G_c=7$ mM, and when the glucose concentration drops below the tipping point on the decreasing part at $G_c=6.6$ mM. The other vertical dashed lines (second and fourth from left) mark approximately the activation and deactivation intervals. We can see that the activation interval occurs well before the cells are fully synchronized and that deactivation interval starts when the cells are desynchronized. 

Next, we map the collective response of \bc{ }{s} to stimuli onto a correlation network where the links between the nodes are determined by the activity and synchronization of cells. The way that functional networks representing islets are usually constructed from calcium dynamics in \bc{ }{s} involves two steps. First, correlations $c_{ij}(t)$ between short segments with length $T$ of calcium signals at time $t$ for all pairs $(i,j)$ of cells are obtained from recorded time series of calcium activity, and then these correlations are compared to some chosen threshold value $c_0$. If correlation of a certain cell pair $(i,j)$ exceeds the threshold value, $c_{ij}(t) > c_0$, the link $(i,j)$ is placed in the network between the nodes representing this cell pair. The choice of the threshold value in such correlation or functional graphs is crucial as it determines the network density and other network properties.

Here, we keep the total number of links $L=N\langle k\rangle/2$, where $\langle k\rangle$ is the mean degree, fixed~\cite{korosak2021} and for particular network realization we select top $L$ links from link list sorted by correlation coefficient. Here we use $\langle k\rangle=10$ throughout this paper. In such correlation network construction we then observe the changes in the linkage arrangment between the nodes as a function of time. These structural changes are reflected in some global network properties such as the largest ($\lambda_1$) and the second largest ($\lambda_2$) eigenvalues of the adjacency matrix, the clustering coefficient ($C$), assortativity ($r$), and the eigenratio $\lambda_N/\lambda_2$ between the largest and the second smallest eigenvalues of the network Laplacian matrix. The time dependence of these network properties are displayed in Figures 2 and 3. 


As glucose concentration approaches the critical tipping point, a sharp increase is observed in both the largest eigenvalue, $\lambda_1$, and the clustering coefficient, $C$, indicating a significant reconfiguration of the \bc{ }{s} network. This reconfiguration suggests a structural transition in the network, despite the number of connections remaining constant. The increase in $\lambda_1$ implies that certain nodes (cells) in the network might be gaining greater influence measured by their number of links, which is captured by the bounds set by the Perron-Frobenius theorem: $\langle k \rangle < \lambda_1 < k_{max}$, where $\langle k \rangle$ is the average degree and $k_{max}$ is the maximum degree of the network. More tight bound for $\lambda_1$ in some random networks is $\max{(\sqrt{k_{max}}, \langle k^2\rangle/\langle k\rangle)}$~\cite{chung2003spectra}, while increasing the variance of interaction strength also leads to increasing of spectral radius or $\lambda_1$~\cite{rajan2006eigenvalue}. The network is not merely random but is reorganizing into a more coordinated, hierarchical structure as glucose crosses the tipping point. The observed changes align with the experimental calcium signaling data, where the \bc{ }{s} shift from an uncoordinated state to a highly synchronized collective response. 

Particularly interesting is the behavior of the second largest eigenvalue $\lambda_2$ near the tipping points (Figure 2) where the transient sharp increase is observed. While the largest eigenvalue $\lambda_1$ follows the time evolution of average correlation and synchronization between \bc{ }{s}, the changes in the second largest eigenvalue $\lambda_2$ occur before the onset of synchronization indicating the structural changes in the network that precede the collective response of \bc{ }{s}. This is similar to the $\lambda_2$ behavior observed in correlation matrices of stock market evolution~\cite{buccheri2013evolution} where the spikes in $\lambda_2$ are associated with the changes in correlation between particular stocks and all others. In the context of \bc{ }{s} activity, the observed changes in $\lambda_2$ suggest that the network is reorganizing its structure so that the onset of synchronization starts with particular cells or small clusters of cells.  

\begin{figure}
    \centering
    \includegraphics[width=0.99\linewidth]{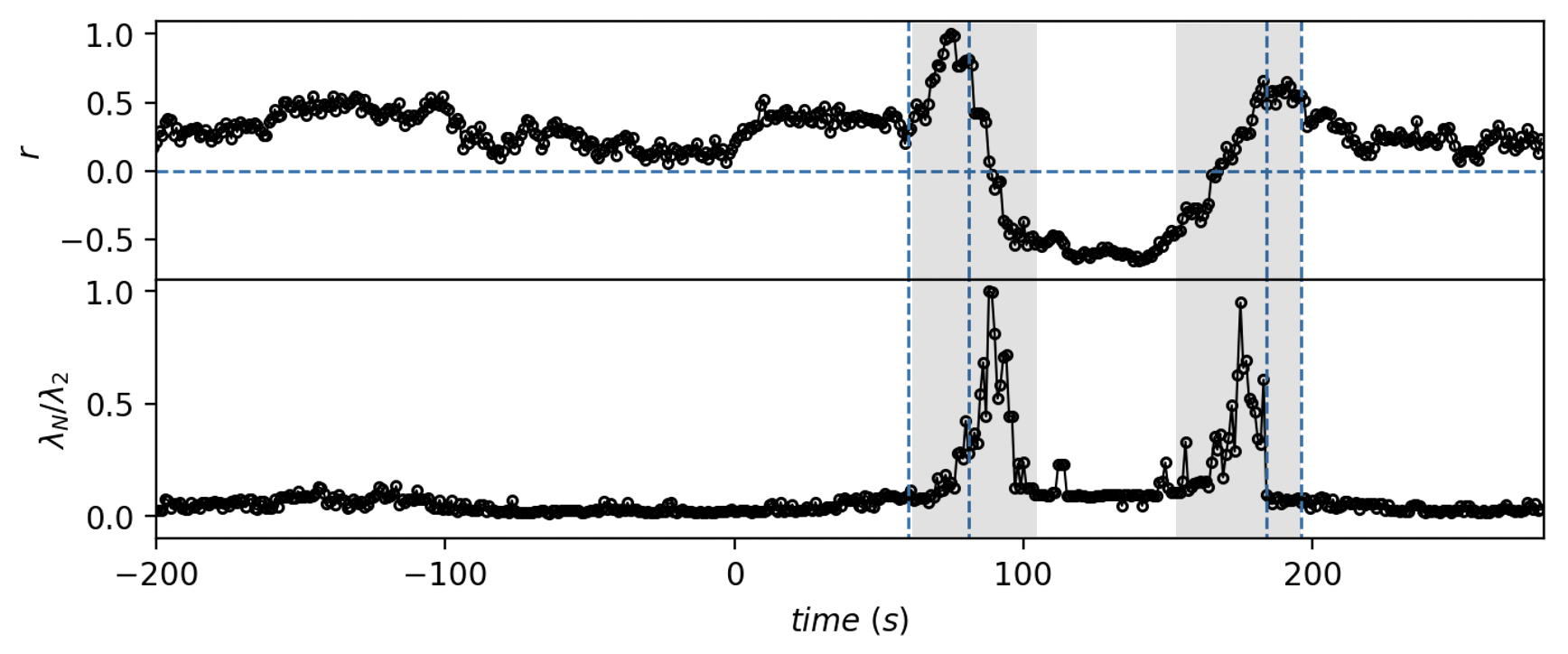}
    \caption{\label{fig:fig3} {\bf Assortativity and Laplacian eigenvalues ratio.} Assortativity coefficient $r$ (top)and ratio of the largest and the second smallest eigenvalues of the network Laplacian $\lambda_N/\lambda_2$ (bottom) as a function of time. The network is constructed with fixed number of links and computations are performed on the largest connected component of the network at each particular time.
    }
\end{figure}

The changes in the network structure during the phase transition are further reflected in the behavior of the assortativity coefficient $r$ and the ratio of the largest to the second smallest eigenvalue of the network Laplacian, $\lambda_N/\lambda_2$, as shown in Figure 3. As the glucose concentration increases and crosses the tipping point, the assortativity coefficient $r$, which quantifies the preference for nodes to link with (dis)similar degree nodes, exhibits a transient increase before the network becomes disassortative $r<0$. This suggests a shift towards a network organizaton where high-degree \bc{ }{s} preferentially connect with low-degree cells. 

Simultaneously, the ratio $\lambda_N/\lambda_2$, representing the spectral gap of the network Laplacian, increases sharply around the tipping points. A larger spectral gap implies that the network ability to propagate signals or synchronize its components is enhanced, which is consistent with the observed transition to a correlated calcium signaling state. This ratio, therefore, serves as a sensitive indicator of the network structural and dynamic changes during the phase transition.

In combination with the results presented in Figure 2, where the largest eigenvalue $\lambda_1$ and clustering coefficient $C$ increase dramatically at the tipping point, these changes highlight a fundamental reorganization of the network topology. The network transitions from a more random, uncoordinated configuration to one that is highly organized and structured. This reorganization correlates with the abrupt increase in the mean calcium signal, demonstrating that the underlying network dynamics are tightly coupled to the observed biological response. 

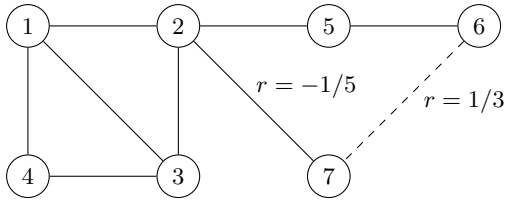
\begin{figure}[ht]
    \centering
    \begin{tikzpicture}[scale=1, transform shape]
        \node[circle, draw] (1) at (0,0) {1};
        \node[circle, draw] (2) at (2,0) {2};
        \node[circle, draw] (3) at (2,-2) {3};
        \node[circle, draw] (4) at (0,-2) {4};
        \node[circle, draw] (5) at (4,0) {5};
        \node[circle, draw] (6) at (6,0) {6};
        \node[circle, draw] (7) at (4,-2) {7};

        \draw (1) -- (2);
        \draw (2) -- (3);
        \draw (3) -- (4);
        \draw (4) -- (1);
        \draw (2) -- (5);
        \draw (5) -- (6);
        \draw[dashed] (6) -- (7);
        \draw (1) -- (3);
        \draw (2) -- (7);
        \draw (3.7,-.8) node {$r=-1/5$};
        \draw (5.8,-1) node {$r=1/3$};
    \end{tikzpicture}
    \caption{{\bf A small graph example.} In a small and simple graph relinking from (6,7) to (2,7), changes the network structure from assortative to disassortative, from $r=1/3$ to $r=-1/5$. This link switch also changes the largest eigenvalue of the adjacencny matrix and the maximal degree of the graph.}
    \label{fig:graph}
\end{figure}

In Figure 4, we illustrate how switching a single link in a simple graph can induce a transition from an assortative to a disassortative network structure. Initially, with the link between nodes 6 and 7, the assortativity coefficient $r = 1/3$ indicates that nodes preferentially connect to others with similar degrees. However, when this link is switched from (6,7) to (2,7), the assortativity changes to $r = -1/5$, indicating a shift to a disassortative structure, where nodes with different degrees become more likely to connect. This link change also affects other network properties, such as the largest eigenvalue of the adjacency matrix and the maximal degree, highlighting how a seemingly minor alteration can significantly impact the network’s overall topology.
This behavior in the simple graph hint towards a key principle in larger networks: small structural changes, such as the rewiring or relinking of a few critical connections, can induce significant transitions in the network’s properties at the critical point.

\begin{figure}
\centering
\includegraphics[width=0.75\linewidth]{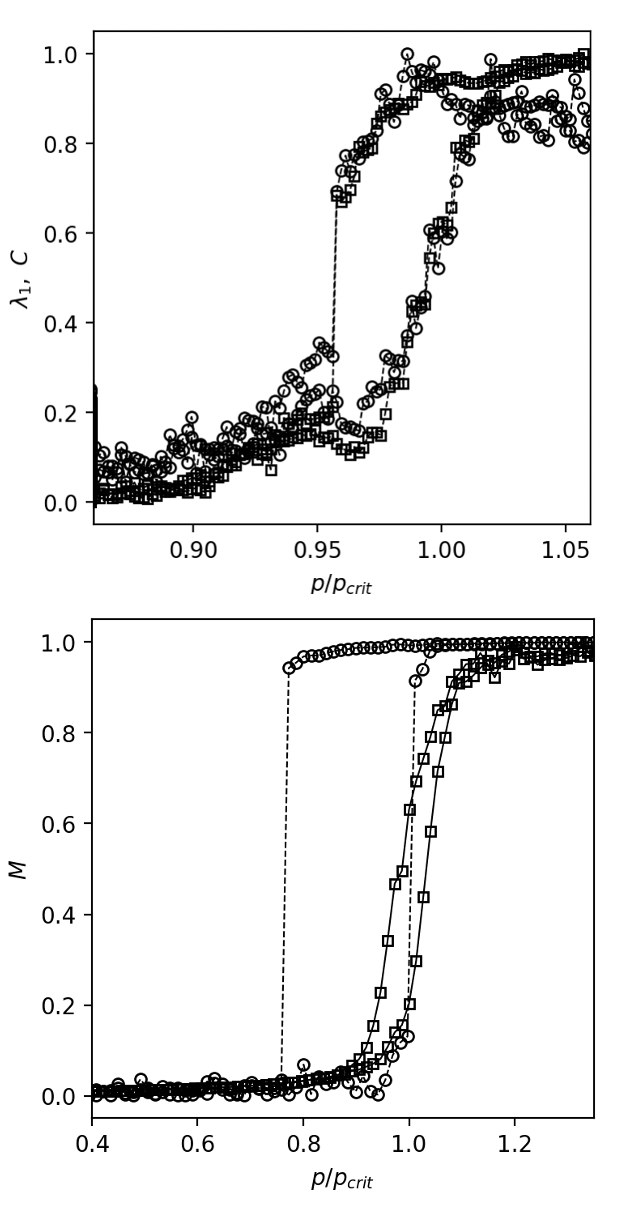}
\caption{\label{fig:fig4} {\bf Phase transition and hysteresis.} Upper panel: Largest eigenvalue of correlation matrix (cirles), clustering of the correlation network (squares) with fixed number of links. Lower panel: Average spin (magnetization) in microscopic model with spins and explosive percolation process 
with $N=10^3$, $k=10$, $q=0.1$ (squares) and $q=0.6$ (circles).
}
\end{figure}

In Figure 5 (upper panel) we plotted $\lambda_1$ and $C$ as a function of the control parameter (the glucose concentration) through the increasing and decreasing part. The hysteresis loop that is clearly visible in both $\lambda_1$ and $C$ as the glucose concentration crosses the tipping point is a hallmark of a first-order phase transition, indicating that the transition from an uncoordinated to a coordinated state of \bc{ }{s} activity is abrupt and irreversible. 

To model the observed empirical findings we introduce a \bc{ }{} network that combines two processes: (1) the link selection process that describes initial intracellular calcium activity occuring in separate, small clusters of \bc{ }{s}, and (2) the coordination of stochastic discrete \bc{ }{s} states that describes the synchronization process accross the entire islet as smaller cell clusters coalesce into large groups building on our previous work~\cite{korosak2021}. We use the explosive percolation mechanims coupled with Ising spins, similarly as in~\cite{angst2012explosive}. We use a random regular network $G(N,k)$ with $N$ nodes (spins, $s_i$) representing \bc{ }{s} which each can be in spin up or spin down state $s_i\in\{1,-1\}$, and where each has $k$ neighbors. 

Initially, the nodes have randomly assigned spins. In each computational step we gradually increase the parameter $p$ that sets the number of links $L=pN$. We start with an empty network successively placing links that results from competition percolation process~\cite{achlioptas2009explosive, nagler2011impact} with three potential links. The three potential links in each step are sampled from a set of viable links composed of all links from $G(N,k)$ that have spins pointing in the same direction at each end. The link that is placed in the network in each step is the one that minimizes the product of sizes of two clusters that the link connects. Finally, each spin in newly connected clusters is assigned the same, but randomly chosen direction with probability $1-q$. If $q=0$ we have complete cluster flipping of spins as in Swendsen-Wang Ising dynamics~\cite{swendsen1987nonuniversal}, while if $q=1$ all spins in the two newly connected clusters are randomly oriented. After each completed computational step we compute the size of the largest connected component and the spin orientation $M=|\sum_i s_i|$. The lower panel of Figure 5 shows the collective spin orientation $M$ computed for two different values of $q$ as a function of $p$. For low values of $q$ we observe a sharp increase in $M$ and hysteresis in dependence on $p$. Also, as $q$ increases the width of the hysteresis decreases and vanishes when $p$ and $q$ reach their critical values in critical point $p=p_c$, $q=q_c$ leading to a second order phase transition. 

Changing parameters $p$ and $q$ allows us to explore the phase space of the model. For $q<<1$ we typically observe a hysteresis in largest component size and magnetization dependence on $p$. As we slowly increase the parameter $p$ we can follow the building of largest component and observe a phase transition when $p$ reaches a critical value. If we start from a connected state of the network and slowly decrease $p$ the largest component vanishes at a different (lower) critical $p$. Simirarly, we observe a hysteresis if we change the parameter $q$ from $q=0$ to $q=1$ and back at some fixed value of parameter $p$. The first example corresponds to correlation network construction where the links are added as the correlation between the nodes increases and exceeds the threshold
value (fixed $q$ in this case). The second example corresponds to the case where we fix the number of links in the network and the network structure changes as the correlation between the nodes increases (in this case $p$ is fixed).  


A simple and highly stylized model offers additional insights into the nature of the phase transition in \bc{ }{s} activity. Let $x$ be the fraction of nonsynchronized cells in the population of cells that are active when the glucose concentration crosses the tipping point. $x$ is therefore the order parameter equal to $x=0$ for fully synchronized and $x=1$ for fully nonsynchronized state. The rate of change is $dx/dt = x(1-x)-Px/(Q+x)$, where the activation (first term) is described as simple logistic growth. The synchronization is described by the second term with two parameters $P$ and $Q$. $P$ can be understood as a measure of the efficiency of communication between cells, with higher $P$ values indicating a more interconnected network, allowing for quicker synchronization due to an increased number of communication links. On the other hand, $Q$ represents a threshold or saturation parameter, which indicates how the synchronization process slows down. In other words, $P$ increases the probability of synchronization through effective communication, while $Q$ introduces a probability that saturation effects will slow down synchronization as the system approaches full synchronization. In this sense $P$ and $Q$ play a role similar to $p$ and $q$ in the microscopic model.

Due to the simple structure, the steady state solution of the model $dx/dt=0$ is analytically tractable~\cite{bose2019bifurcation} and leads to the phase space where two spinodals $P=Q$ and $P=(1+Q)^2/4$ separate the bistability region from stable nonsynchronized and synchronized states. Traversing the phase space with increasing and decreasing $P$ at fixed $Q$ the system undergoes a first order phase transition from $x=1$ to $x=0$ and back with hysteresis. The width of the hystersis is determined by the difference between the two tipping points and equals $\Delta H = (1-Q)^2/4$. The hysteresis vanishes at the critical point $Q_c=1$, $P_c=1$ where the system undergoes a second order phase transition. 

Similarly, we can make some estimates to better understand how the correlation structure of the \bc{ }{} collectives influences the likelihood of connections forming between nodes in the network modeling them. 
Let the strength of a node $k_i(t)$ be the sum of all correlations with all other nodes $k_i(t) = \sum_{i \neq j} c_{ij}(t)/(N-1)$.
We describe the interaction of two nodes with $m_{ij}(t) = k_i(t)k_j(t)$ and define the probability $h(t)$ that any two nodes in the network are linked characterising the islet network state at time $t$ as the interaction between the nodes averaged over all node pairs $h(t) = \langle m_{ij}\rangle$. This approach is similar to the configuration network model where given the node degree sequence the edge probability between nodes is defined as proportional to the product of node degrees. In the large network limit $N >> 1$ or when variance of the correlation distribution is sufficiently small we simply have 
$h = c_m^2$, where $c_m$ is the mean correlation coefficient at time $c_m = \sum_i k_i/N$. The probability to connect two nodes in the this random network model is therefore equal to square of mean correlation. Since the critical probability for the rise of the giant component in a random network with $N$ nodes is $h_c = 1/N$, we have a simple measure for the giant component to appear: the mean correlations in the system must exceed $c_m > 1/\sqrt{N}$. 

Our results provide compelling evidence that \bc{ }{s} undergo a first-order phase transition in response to glucose stimulation, with distinct hysteresis. This behavior suggests that pancreatic islets function as tipping elements, where small changes in glucose can lead to abrupt shifts in \bc{ }{s} activity.

These findings have significant implications for our understanding of insulin secretion dynamics. The sharp transitions observed in \bc{ }{s} activity may explain the biphasic nature of insulin release, where an initial rapid response is followed by a slower, more sustained phase. Additionally, the presence of hysteresis suggests that \bc{ }{s} collectives may retain memory of previous glucose levels, potentially contributing to the oscillatory nature of insulin secretion. Each islet functions as a localized tipping point, where small changes in glucose concentration trigger a collective, synchronized response from  \bc{ }{s}, leading to insulin secretion. However, islets do not operate in isolation; they are interconnected through perfusion, innervation, and diffusion networks within the pancreas. These connections mean that the behavior of one islet can influence neighboring islets, creating a cascade effect where synchronized activity in one islet can propagate through the system, ensures a rapid, robust insulin release, but disruptions in these connections could lead to impaired responses, as seen in conditions like diabetes.

In conclusion, we demonstrate that \bc{ }{s} in pancreatic islets undergo a first-order phase transition, with distinct hysteresis, in response to changing glucose levels. These findings provide new insights into the dynamics of insulin secretion and highlight the role of islets as critical regulators of metabolic homeostasis. Our network model offers a promising framework for further exploration of phase transitions in biological systems.

\begin{center}
--\,--\,--\,--\,--
\end{center}
\vspace{1mm}
\noindent\textbf{Acknowledgements.} MSR received financial support from by the Austrian Science Fund / Fonds zur F{\"o}rderung der Wissenschaftlichen Forschung (bilateral grants I3562-B27 and I4319-B30) and NIH (R01DK127236). DK, AS and MSR received financial support from the Slovenian Research Agency (research core funding program P3-0396 and projects N3-0048, N3-0133 and N3-9289). DK and BP were also supported from the Slovenian Research Agency project no.J7-3156.

\noindent\textbf{Code and data availability.} The data and the code for implementation of the microscopic model, the network construction and data analysis is available at \url{https://github.com/deankorosak/critical_transitions/} with additional data avaliable at \url{https://doi.org/10.6084/m9.figshare.25507123}.

\noindent\textbf{Author contributions.} All authors contributed substantially to all aspects of the study.

\noindent\textbf{Conflict of interest.} The authors declare no conflict of interest, financial or otherwise.

\bibliography{refs}{}
\bibliographystyle{apsrev4-1}

\end{document}